\documentclass[prd,onecolumn,preprintnumbers,nofootinbib]{revtex4}
\usepackage{graphicx}

\usepackage[plainpages=false, colorlinks=true, anchorcolor=blue, linkcolor=blue, citecolor=blue, bookmarks=false]{hyperref}
\newcommand{\rthis}[1]{\textcolor{black}{#1}}
\usepackage{amsfonts,amsmath,amssymb}
\usepackage{graphicx}
\usepackage{pdfpages}
\usepackage{booktabs}
\usepackage{hhline}
\usepackage[font=footnotesize]{caption}
\usepackage{multirow}
\usepackage{url}
\pdfoutput=1

\begin{document}
\pdfoutput=1
\title {Bayesian Model Comparison of $R_h=ct$ versus $\Lambda$CDM using HII galaxy Hubble diagram}
\author{Yuva  Himanshu \surname{Pallam}$^1$} \altaffiliation{E-mail:f20220962@hyderabad.bits-pilani.ac.in}
\author{Shantanu \surname{Desai}$^2$ }
\altaffiliation{E-mail: shntn05@gmail.com}


\begin{abstract}
We complement a recent analysis  comparing $R_\mathrm{h}=ct$ 
with $\Lambda$CDM/$w$CDM  using HII galaxies and giant extragalactic HII regions, by carrying out Bayesian model comparison. For this purpose, we calculate the Bayes factors for $R_\mathrm{h}=ct$ 
compared to flat $\Lambda$CDM/$w$CDM using the same dataset. When we use uniform priors on cosmological parameters,  we find that the Bayes factors are close to 1, implying that $R_\mathrm{h}=ct$ is equally favored compared to $\Lambda$CDM/$w$CDM. However, when we use normal priors on cosmological parameters based on Planck  cosmology, we find that $R_\mathrm{h}=ct$ is strongly favored over flat $\Lambda$CDM, while $R_\mathrm{h}=ct$ is marginally favored  over flat $w$CDM.
\end{abstract}

\affiliation{$^{1}$Department of Physics, BITS Pilani - Hyderabad Campus, Jawahar Nagar, Shameerpet Mandal, Hyderabad,  Telangana-500078, India}
\affiliation{$^{2}$Department  of Physics, IIT Hyderabad,  Kandi, Telangana-502284, India}
\maketitle
\section{Introduction}
In a recent work, \citet{Melia} (WM25 hereafter) used a sample of sources consisting of HII galaxies (HIIGx) and giant extragalactic HII regions (GEHR)  for cosmological analyses. These HIIGx and GEHRs are massive and compact starburst galaxies, where most of the luminosity is emanated  from rapidly forming stars surrounded by ionized hydrogen~\cite{Melnick87}. For these galaxies, the H$\beta$ luminosity has been shown to be strongly correlated with the ionized gas velocity dispersion with very small scatter~\cite{Terlevich81}.
Therefore, these galaxies have turned out to be very good standard candles and have been used as cosmic distance indicators and probes of cosmic expansion history~\cite{Melnick00,Siegel05,Mania2011,CaoRatra20,CaoRyanKhadka,Cao21,Cao22,Cao23,Johnson,Mehrabi21,Gao24}.

WM25  analyzed a compilation of 231 HIIGx and GEHRs in the redshift range from $z \approx 0$ to $z \approx 7.5$ to carry out a comparative test of three different cosmologies: flat-$\Lambda$CDM, $w$CDM and $R_\mathrm{h}=ct$. For this purpose,   Bayesian  Information Criterion-based model comparison was used to rank these models~\cite{Liddle07}.
WM25 showed that $R_\mathrm{h}=ct$ is favored over both flat-$\Lambda$CDM and $w$CDM. The difference in BIC between $R_\mathrm{h}=ct$ and  flat-$\Lambda$CDM  as well as  $w$CDM was found to be about 5.0 and 9.5, respectively.  These correspond to ``positive'' and ``strong'' evidence for 
$R_\mathrm{h}=ct$ over $\Lambda$CDM and $w$CDM, respectively,  using the qualitative strength of evidence rules~\cite{Krishak}.
WM25  further redid this model selection analysis by adding an  unknown intrinsic scatter in the regression relation. With this additional parameter,  $R_\mathrm{h}=ct$ is no longer  favored as before. With this change, WM25 found that  $R_\mathrm{h}=ct$ and flat-$\Lambda$CDM are equally favored with  $\Delta BIC<1$, while  $R_\mathrm{h}=ct$  is positively  favored over $w$CDM with $\Delta BIC \sim 5.0$. However, after the addition of an intrinsic scatter, there is a $2.5\sigma$ discrepancy between the inferred $\Omega_\mathrm{m}$ value as compared to the Planck  cosmology analysis~\cite{Planck20}.

In this work, we complement the analysis in WM25 by using Bayesian model comparison to choose the best cosmological model among the three cosmological models following the same methodology. Bayesian model comparison does not make use of the best-fit values of the parameters obtained from maximizing the likelihood,  and is much more  robust than information theory-based model comparison techniques~\citep{Trotta,Weller,Sanjib,Krishak,Liddle07}. BIC also assumes that the dataset is independent and identically distributed, whereas no such assumptions are made in Bayesian Model selection~\citep{Liddle07}. This is a followup to our previous works, where we compared $R_\mathrm{h}=ct$  and $\Lambda$CDM using cosmic chronometers~\citep{Haveesh}, galaxy cluster gas mass fractions~\citep{Kunj1}, and Baryon Acoustic Oscillations (BAO)~\citep{Kunj2}. Other successes and failures of $R_\mathrm{h}=ct$  have been recently reviewed in WM25 and one of our previous works~\cite{Kunj2}, respectively, including in references therein.
Finally, we note that currently the best-fit constraints on the concordance $\Lambda$CDM (or generalized dark energy models such as $w$CDM) mainly come from CMB, BAO, and Type Ia SN. We have recently shown that  the DESI BAO results  decisively favour  $\Lambda$CDM over $R_\mathrm{h}=ct$~\citep{Kunj2}. However, this result hinges upon only one data point, viz. the Lyman-$\alpha$ QSO measurement at $z=2.33$~\citep{DESIDR1}. All other BAO measurements in DESI DRI are consistent with both $\Lambda$CDM and $R_\mathrm{h}=ct$. If this measurement gets revised or in case the uncertainties were underestimated, it could change the results. Therefore, it would be beneficial to  use additional  observational probes to test the $R_\mathrm{h}=ct$ cosmological model.
Although the CMB provides a robust measure of the angular diameter distance, it cannot be directly  used to probe  $R_\mathrm{h}=ct$ and other power law cosmological models, because the redshift at which the CMB decouples is not model independent~\citep{Shafer15}. Type Ia Supernovae have previously been used to probe the efficacy of $R_\mathrm{h}=ct$ compared to $\Lambda$CDM in a number of works~\citep{Shafer15,WeiMelia15}. We shall defer a similar analysis using the latest SNe datasets to a future work.

This manuscript is structured as follows. Our analysis is described in Sect.~\ref{sec:analysis} and results  in Sect.~\ref{sec:results}. We conclude in Sect.~\ref{sec:conclusions}.

\section{Analysis and Results}
\label{sec:analysis}
We provide a brief recap of the regression analysis  done in WM25. We use the same notation as that used in WM25. Note that we also independently carry out the same regression calculation and BIC-based model comparison in Appendix A.
  
The dataset used consists of a sample of 231 sources comprised of 
 36 GEHRs and nearby HIIGx~\citep{Chavez25} with independently measured distance moduli~\citep{fernandez}, joint local and high-$z$ sample of 181 HIIGx~\citep{gonzalez}, and 14 newly discovered HIIGx samples~\citep{Llerena23,DeGraaf}.
For our Bayesian inference, similar to WM25,  we construct the following likelihood, modeled as a sum of separate likelihoods for the HIIGx and anchor samples:
\begin{equation}
\mathrm{ln}(\mathcal{L}_\mathrm{tot}) = \mathrm{ln}(\mathcal{L}_\mathrm{HIIGx}) + \mathrm{ln}(\mathcal{L}_\mathrm{anchor}),
\label{eq:L}
\end{equation}
where $\mathcal{L}_\mathrm{HIIGx}$ is given by:
\begin{equation}
\mathcal{L}_\mathrm{HIIGx} = \prod_{i}^{195}\frac{1}{\sqrt{2\pi}\,\epsilon_{\mathrm{HIIGx,i}}}\times \exp\left[-\frac{(\mu_{\mathrm{obs},i}-\mu_{\mathrm{th}}(z_i))^2}{2\epsilon^2_{\mathrm{HIIGx},i}}\right],
\label{eq:L1}
\end{equation}
where
\begin{equation}
\epsilon^2_{\mathrm{HIIGx},i}=\sigma^2_{\mu_{\mathrm{obs}},i}+\left[\frac{5\,\sigma_{D^{\mathrm{th}}_L,i}}{\mathrm{ln}\,10\,D^{\mathrm{th}}_L(z_i)}\right]^2.
\label{eq:epshii}    
\end{equation}
 $\mathcal{L}_\mathrm{anchor}$ in Eq.~\ref{eq:L} is given by: 
\begin{equation}
\mathcal{L}_\mathrm{anchor} = \prod_{i}^{36}\frac{1}{\sqrt{2\pi}\,\epsilon_{\mathrm{anchor},i}}\times \exp\left[-\frac{(\mathrm{log}\,L(\mathrm{H\beta})_i-\beta \, \mathrm{log} \,\sigma_i-\alpha)^2}{2\epsilon^2_{\mathrm{anchor},i}}\right],
\label{eq:L2}
\end{equation}
where $\epsilon_{\mathrm{anchor},i}$ is given by:
\begin{equation}
\epsilon_{\mathrm{anchor},i}^2=\sigma_{\mathrm{log}\,L,i}^2+\beta^2\,\sigma_{\mathrm{log}\,\sigma,i}^2
\label{eq:epsanchor}
\end{equation}

The  observed distance modulus  $\mu_{\mathrm{obs}}$ in Eq.~\ref{eq:L1} is given by:
\begin{equation}
    \mu_{\mathrm{obs}}=2.5\,[\beta\, \mathrm{log}\,\sigma + \alpha-\mathrm{log}\,F(\mathrm{H}\beta)]-100.2
\end{equation}
The corresponding error in $\mu_{\mathrm{obs}}$   ($\sigma_{\mu_{\mathrm{obs}}}$) can be obtained using error propagation and is given by: 
\begin{equation}
    \sigma_{\mu_{\mathrm{obs}}}=2.5\,(\beta^2\, \sigma^2_{\mathrm{log}\,\sigma} + \sigma^2_{\mathrm{log}\,F})^{1/2} .
\end{equation}
\rthis{We should point out that one assumption in using the HII galaxies as standard candles to probe different expansion histories, is that there is no evolution in the $L-\sigma$ scaling relation as a function of  redshift. However,  evidence for such  an evolution in the $L-\sigma$ relation  for the dataset of 181 HIIGx galaxies~\cite{gonzalez} between the low-redshift sample (107 galaxies, $z=0.0088-0.164$) and the high-redshift sample (74 galaxies, $z=0.634-2.545$) has been found~\cite{Cao24}. Therefore,  the complete dataset  may not be standardizable in a uniform way. This is one caveat in our analysis and also in WM25. }

The theoretical distance modulus $\mu_{\mathrm{th}}$ is related to the luminosity distance $D_L(z)$ at redshift ($z$) using the following expression:
\begin{equation}
\mu_{\mathrm{th}}\equiv 5\,\mathrm{log}\left[\frac{D_L(z)}{\mathrm{Mpc}}\right]+25
\end{equation}
The luminosity distance for generalised $w$CDM with nonzero curvature is given by:
\begin{equation}
\begin{split}
    D_L^{w\mathrm{CDM}}(z) &= 
    \frac{c}{H_0}\frac{(1+z)}{\sqrt{|\Omega_k|}}
    \, \mathrm{sinn}\!\Bigg\{
    \sqrt{|\Omega_k|} \times \\
    & \quad \int_0^z \frac{dz}{%
    \sqrt{\Omega_\mathrm{m}(1+z)^3 
    + \Omega_k(1+z)^2 
    + \Omega_{\mathrm{de}}(1+z)^{3(1+w_{\mathrm{de}})}}}
    \Bigg\}.
\end{split}
\end{equation}
While comparing $R_\mathrm{h}=ct$ and  $w$CDM, we also consider $\Omega_k=0$, similar to WM25. The latest cosmological results using late-universe probes are also consistent with a flat universe~\citep{Barua}.
The luminosity distance for generalized $\Lambda$CDM with nonzero curvature
is given by:

\begin{equation}
\begin{split}
    D_L^{\Lambda \mathrm{CDM}}(z) &=
    \frac{c}{H_0}\frac{(1+z)}{\sqrt{|\Omega_k|}}
    \, \mathrm{sinn}\!\Bigg\{
    \sqrt{|\Omega_k|} \times \\
    & \quad \int_0^z \frac{dz}{%
    \sqrt{\Omega_\mathrm{m}(1+z)^3 
    + \Omega_k(1+z)^2 
    + \Omega_\Lambda}}
    \Bigg\}.
\end{split}
\end{equation}

The flat-$\Lambda$CDM model is a special case with $\Omega_k=0$  and $\Omega_\Lambda = 1-\Omega_\mathrm{m}$. 
The corresponding expression for $D_L$ in  $R_\mathrm{h}=ct$ cosmology is given by~\citep{Melia12}
\begin{equation}
    D_L^{R_\mathrm{h}=ct}(z) = \frac{c}{H_0}(1+z)\,\mathrm{ln}(1+z).
\end{equation}
We evaluated the  luminosity distance for   $\Lambda$CDM and $w$CDM cosmologies using {\tt astropy}~\citep{astropy}.

For our analyses we considered two sets of priors.
We first used uniform priors on cosmological parameters given by $H_{\mathrm{0}} \in \mathcal{U}$ (60,110) km/sec/Mpc, $\Omega_{\mathrm{m}} \in \mathcal{U} (0,1),w_{\mathrm{de}} \in \mathcal{U} (-3,0)$.
We then also chose Gaussian priors on the cosmological parameters
using the best-fit values from Planck cosmology analysis,  given by
 $H_0 \in  \mathcal{N} (67.4,0.5)$ km/sec/Mpc, and $\Omega_{m} \in \mathcal{N} (0.315,0.007)$~\citep{Planck20}. \rthis {We note however that the choice of Planck derived priors is not a model-neutral comparison against $R_\mathrm{h}=ct$, since these are obtained  within a specific early-universe framework and $\Lambda$CDM parameterization. However they serve as an additional comparison to test the viability of $R_\mathrm{h}=ct$.}
For the priors on astrophysical parameters  we used, $\alpha \in \mathcal{U} (33,35), \beta \in \mathcal{U} (4,5)$, which are same in both the cases. When we consider intrinsic scatter, we use uniform prior   given by $\mathcal{U} (0,1)$.

\subsection{Accounting for intrinsic scatter in observables}
WM25 also carried out an augmented analysis to account for an unknown intrinsic scatter ($\sigma_{int}$) in the correlation between $L(\mathrm{H\beta})$ and $\sigma$. For this purpose, $\epsilon_{\mathrm{HIIGx}}$ and  $\epsilon_{\mathrm{anchor}}$ are modified as follows:
\begin{equation}
\epsilon^2_{\mathrm{HIIGx},i}=6.25\left(\sigma_{\mathrm{int}}^2+\beta^2\sigma^2_{\mathrm{log}\,\sigma,i}+\sigma_{\mathrm{log}\,F,i}^2\right)+\left[\frac{5\sigma_{D^{\mathrm{th}}_L,i}}{\mathrm{ln}\,10\,D^{\mathrm{th}}_L(z_i)}\right]^2,
\label{eq:epshiimod}
\end{equation}
\begin{equation}
\epsilon_{\mathrm{anchor},i}^2=\sigma_{\mathrm{int}}^2+\sigma_{\mathrm{log}\,L,i}^2+\beta^2\,\sigma_{\mathrm{log}\,\sigma,i}^2.
\label{eq:epsanchormod}
\end{equation}

In addition to BIC based model selection, WM25 also obtained marginalized credible intervals for each of the free parameters in the aforementioned cosmological models. We also redid our analysis after accounting for intrinsic scatter and using the same sets of priors as before.
 
\section{Results of Bayesian model comparison}
\label{sec:results}
We now implement the Bayesian model comparison between $R_\mathrm{h}=ct$ and flat-$\Lambda$CDM/$w$CDM cosmological models. Before presenting our results,  we provide a very brief primer on Bayesian model comparison and defer the reader to  various reviews for further details~\citep{Trotta,Weller,Sanjib,Krishak}.
To evaluate the significance of a model ($M_2$) as compared to another model ($M_1$), we define  the Bayes factor ($B_{21}$) given by:
\begin{equation}
B_{21}=    \frac{\int P(D|M_2, \theta_2)P(\theta_2|M_2) \, d\theta_2}{\int P(D|M_1, \theta_1)P(\theta_1|M_1) \, d\theta_1} ,  \label{eq:BF}
\end{equation}
where $P(D|M_2,\theta_2)$ is the likelihood for the model $M_2$ for the data $D$, and $P(\theta_2|M_2)$ denotes the prior on the parameter vector $\theta_2$ of the model $M_2$.  The denominator in Eq.~\ref{eq:BF} denotes the same for model $M_1$. For our analysis, the likelihood is specified by Eq.~\ref{eq:L}.
 If $B_{21}$ is greater than one, then the model $M_2$ is preferred over $M_1$ and vice-versa. The significance can be qualitatively assessed using Jeffreys' scale. There are different versions of Jeffreys' scale quoted in literature~\citep{Trotta,Bellido,Krishak,Weller,Kiblinger}. For our analysis, we quote the results from ~\citet{Weller}, since it directly follows the original scale  proposed by Harold Jeffreys~\citep{Jeffreys}. 
 To calculate Bayesian evidence and Bayes factors, we use the nested sampler {\tt Dynesty}~\citep{dynesty}. 
 \rthis{We run {\tt dynesty} using dynamic nested sampling mode. The sampler was run with 1024 initial live points using the multi–ellipsoidal bounding method and uniform sampling of the prior space.  The
  The runs were terminated using the default  stopping criterion,  viz. $d(\ln Z)<0.1$, where $\ln Z$ is the natural log of evidence, and $d(\ln  Z)$ represents the difference in values of $\ln Z$ between successive iterations.  To test the convergence criterion, we reran the codes multiple times to check the stability.}
 We consider model $M_2$ to be $R_\mathrm{h}=ct$ and $M_1$ (which is the null hypothesis) to be the flat-$\Lambda$CDM and  flat-$w$CDM. Therefore, if the Bayes factors are much greater than one, then $R_\mathrm{h}=ct$ is preferred over flat $\Lambda$CDM/$w$CDM. 
 The results for these Bayes factors can be found in Table~\ref{tablemodbf1} without intrinsic scatter, and Table~\ref{tablemodbf2} after accounting for intrinsic scatter in 
 $\epsilon_{\mathrm{anchor}}$ and $\epsilon_{\mathrm{HIIGx}}$.  Note that the total statistical and systematic errors in the calculation of Bayesian evidence are $\mathcal{O}$ (0.01)\% and are not reported.
 
For uniform priors on cosmological parameters, we find that the Bayes factors for $R_\mathrm{h}=ct$ compared to flat-$\Lambda$CDM as well as flat-$w$CDM are close to one, indicating that both models are equally favored based on Jeffreys' scale. For Gaussian priors on cosmological parameters, we find that the Bayes factors for $R_\mathrm{h}=ct$ over $\Lambda$CDM and $w$CDM are around 50 and 6, respectively. This implies ``very strong'' and ``substantial'' evidence for $R_\mathrm{h}=ct$ over   $\Lambda$CDM and $w$CDM, respectively.
Therefore, only for a Gaussian prior on the cosmological parameters obtained from Planck 2020 cosmology, do the results from  Bayesian model comparison  agree  with the conclusions in WM25.  \rthis{However, we note that this conclusion is mainly driven by the tension  between the HIIGx based cosmological constraints and Planck cosmological parameters (cf. Appendix A), rather than  by decisive discrimination from the HIIGx Hubble diagram itself.}

With the inclusion of intrinsic scatter, the results from Bayesian model comparison agree with the BIC-based results in  WM25 for uniform priors. For Gaussan priors, the Bayes factors for $R_\mathrm{h}=ct$ over   $\Lambda$CDM and $w$CDM are equal to 18.5 and 2.5, corresponding to ``strong'' and ``barely worth mentioning'' evidence, respectively.

\begin{table*}
\centering
\renewcommand{\arraystretch}{1.3}
\begin{tabular}{|c|c|c|c|c|c|}
\hline
 & $\ln Z$ ($\Lambda$CDM) & $\ln Z$ ($w$CDM) & $\ln Z$ ($R_\mathrm{h}=ct$) & \textbf{BF}($R_\mathrm{h}=ct/\Lambda$CDM) & \textbf{BF}($R_\mathrm{h}=ct/w$CDM) \\
\hline

Uniform Prior
& $-215.02 \pm 0.08$
& $-215.21 \pm 0.08$
& $-216.35 \pm 0.06$
& $0.27 \pm 0.02$
& $0.32 \pm 0.03$ \\

\hline

Normal Prior
& $-235.62 \pm 0.06$
& $-233.22 \pm 0.07$
& $-231.66 \pm 0.06$
& $50.00 \pm 4.00$ 
& $4.80 \pm 0.40$ \\

\hline
\end{tabular}
\caption{\label{tablemodbf1} Bayes factor ($B_{21}$)  for $R_\mathrm{h}=ct$ as compared to flat  $\Lambda$CDM and $w$CDM. \rthis{The first three columns contain the natural log of the Bayesian evidences along with errors for all three hypotheses.}
\rthis{We find that Bayes factors are $< 1.0$ ($\Lambda$CDM) and $<5.0$ ($w$CDM)} for the case of uniform priors (third column) on $H_0$ and $\Omega_{m}$, which are $\mathcal{U} (60,110)$ and $\mathcal{U}(0,1)$ respectively,  indicating both $w$CDM and $\Lambda$CDM are equally favored compared to $R_\mathrm{h}=ct$. When we use case  normal priors (fourth column) on $H_0$ and $\Omega_{m}$, given by $\mathcal{N} (67.4,0.5)$ and $\mathcal{N} (0.315,0.007)$, the Bayes factors are around 50 and 6, respectively. Consequently, the Jeffreys' scale points to ``very strong'' and ``substantial'' evidence for $R_\mathrm{h}=ct$ over $\Lambda$CDM and $w$CDM, respectively.}
\end{table*}

\begin{table*}
\centering
\renewcommand{\arraystretch}{1.3}
\begin{tabular}{|c|c|c|c|c|c|}
\hline
 & $\ln Z$ ($\Lambda$CDM) & $\ln Z$ ($w$CDM) & $\ln Z$ ($R_\mathrm{h}=ct$) & \textbf{BF}($R_\mathrm{h}=ct/\Lambda$CDM) & \textbf{BF}($R_\mathrm{h}=ct/w$CDM) \\
\hline

Uniform Prior
& $-138.92 \pm 0.09$
& $-139.10 \pm 0.09$
& $-139.65 \pm 0.08$
& $0.48 \pm 0.06$ 
& $0.58 \pm 0.07$ \\

\hline

Normal Prior
& $-168.96 \pm 0.07$
& $-166.97 \pm 0.08$
& $-166.04 \pm 0.07$
& $18.60 \pm 1.90$
& $2.54 \pm 0.28$ \\

\hline
\end{tabular}
\caption{\label{tablemodbf2} Bayes factor ($B_{21}$)   for $R_\mathrm{h}=ct$ as compared to flat  $\Lambda$CDM and $w$CDM after accounting for intrinsic scatter in $\epsilon_{\mathrm{anchor}},i$ (cf. Eq.~\ref{eq:epsanchormod}) and $\epsilon_{\mathrm{HIIGx}}$ (cf. Eq.~\ref{eq:epshiimod}) for both uniform and Gaussian priors on cosmological parameters, which are the same as that in Table~\ref{tablemodbf1}. \rthis{The first three columns contain the natural log of the Bayesian evidences along with errors for all three hypotheses.}
Similar to Table~\ref{tablemodbf1}, the Bayes factors are \rthis{$< 1.0$ ($\Lambda$CDM) and $<5.0$ ($w$CDM)} for the case of uniform prior on $H_0$ and $\Omega_{m}$, implying that $w$CDM/$\Lambda$CDM  is equally favored compared to  $R_\mathrm{h}=ct$. For normal priors, the Bayes factors point to  ``strong'' and ``barely worth mentioning'' evidence for $R_\mathrm{h}=ct$ as compared to  $\Lambda$CDM  and $w$CDM, respectively.}
\end{table*}

\section{Conclusions}
\label{sec:conclusions}
In a recent work, WM25 used an updated catalog of HIIGx and GEHRs as standard candles to probe the expansion history of the universe and test the relative efficacy of $R_\mathrm{h}=ct$  compared to $\Lambda$CDM and  $w$CDM. For this purpose, the model comparison was carried out using BIC-based information theory criterion. It was found in that work that  $R_\mathrm{h}=ct$ is strongly favored over $\Lambda$CDM and  $w$CDM, when no additional intrinsic scatter in the observable was considered.

We redid the same analysis in WM25 and performed model selection analysis using Bayesian model comparison, calculating the Bayes factors between the two sets of models. We did our analysis using two sets of priors. We first used uniform priors on the cosmological parameters, viz. $H_0$ and $\Omega_m$,  as well as  normal priors on $H_0$ and $\Omega_m$ based on Planck  2020 Cosmology.
Our results for the same can be found in Table~\ref{tablemodbf1} and  Table~\ref{tablemodbf2}. When we use uniform priors, we find that the Bayes factors for $R_\mathrm{h}=ct$ compared to flat $\Lambda$CDM and flat $w$CDM are close to one, irrespective of whether we consider intrinsic scatter in the regression relation or not. Therefore, for uniform priors on cosmological parameters, both models are equally favored and the results differ with the corresponding BIC based analysis done in WM25. However, when we choose normal priors based on Planck  cosmology, Jeffreys scale indicates  very strong/strong  evidence for  $R_\mathrm{h}=ct$ over flat $\Lambda$CDM, without and with intrinsic scatter, respectively in agreement with WM25.  \rthis{However, this is mainly driven by a tension between the Planck priors  and 
HIIGX only constraints (cf. Fig~\ref{fig:lcdm} in Appendix A).}
Similarly, the Bayes factors shows substantial and barely worth a mention  evidence for $R_\mathrm{h}=ct$ over flat  $w$CDM, without and with intrinsic scatter, respectively. Therefore, the strong evidence for $R_\mathrm{h}=ct$ reported in WM25 is largely an an artifact of the information criterion used, and is confirmed using Bayesian model comparison. 

In the spirit of open science,  our analysis codes and the dataset used have been made publicly available, which can be found at \url{https://github.com/Yuva12345/HII-galaxy-Hubble-diagram}

\section*{Acknowledgements}
YHP has been supported by a Summer Undergraduate Research Exposure (SURE) internship at IIT Hyderabad during the summer of 2025. \rthis{We are grateful to the anonymous referees for several constructive comments on the manuscript.}

\bibliography{main}

\appendix 
\section{Comparison of BIC}
Given the disagreement between the conclusions based on Bayes factor compared to those obtained using BIC in WM25 when using uniform priors on cosmological parameters, we now also reproduce the results of BIC obtained in WM25. 
BIC is given by the following expression~\citep{Liddle07}:
\begin{equation}
\mathrm{BIC}= -2 \ln \mathcal{L}_{max} + p \ln N, 
\end{equation}
where $\mathcal{L}_{max}$ is the maximum value of the likelihood defined in Eq.~\ref{eq:L}; $p$ is the number of free parameters and $N$ is the total number of data points.

\begin{figure*}
    \centering
    \includegraphics[width=0.9\textwidth]{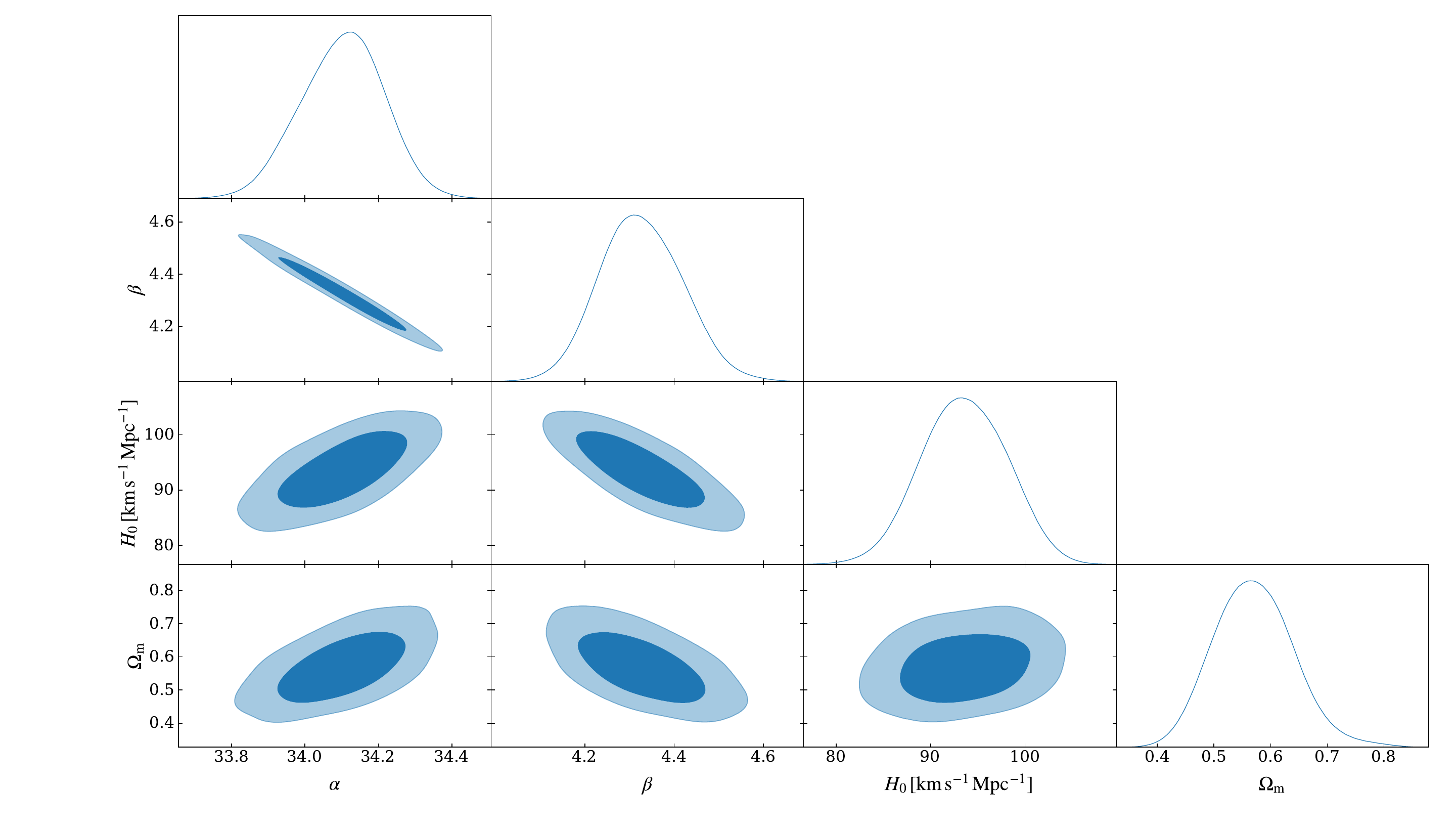}
    \caption{\label{fig:lcdm}
  The marginalized 68\% and 95\% credible intervals for the  parameters $\alpha,\beta,H_0$ and $\Omega_m$ in flat $\Lambda$CDM.}
\end{figure*}

\begin{figure*}
    \centering
    \includegraphics[width=0.9\textwidth]{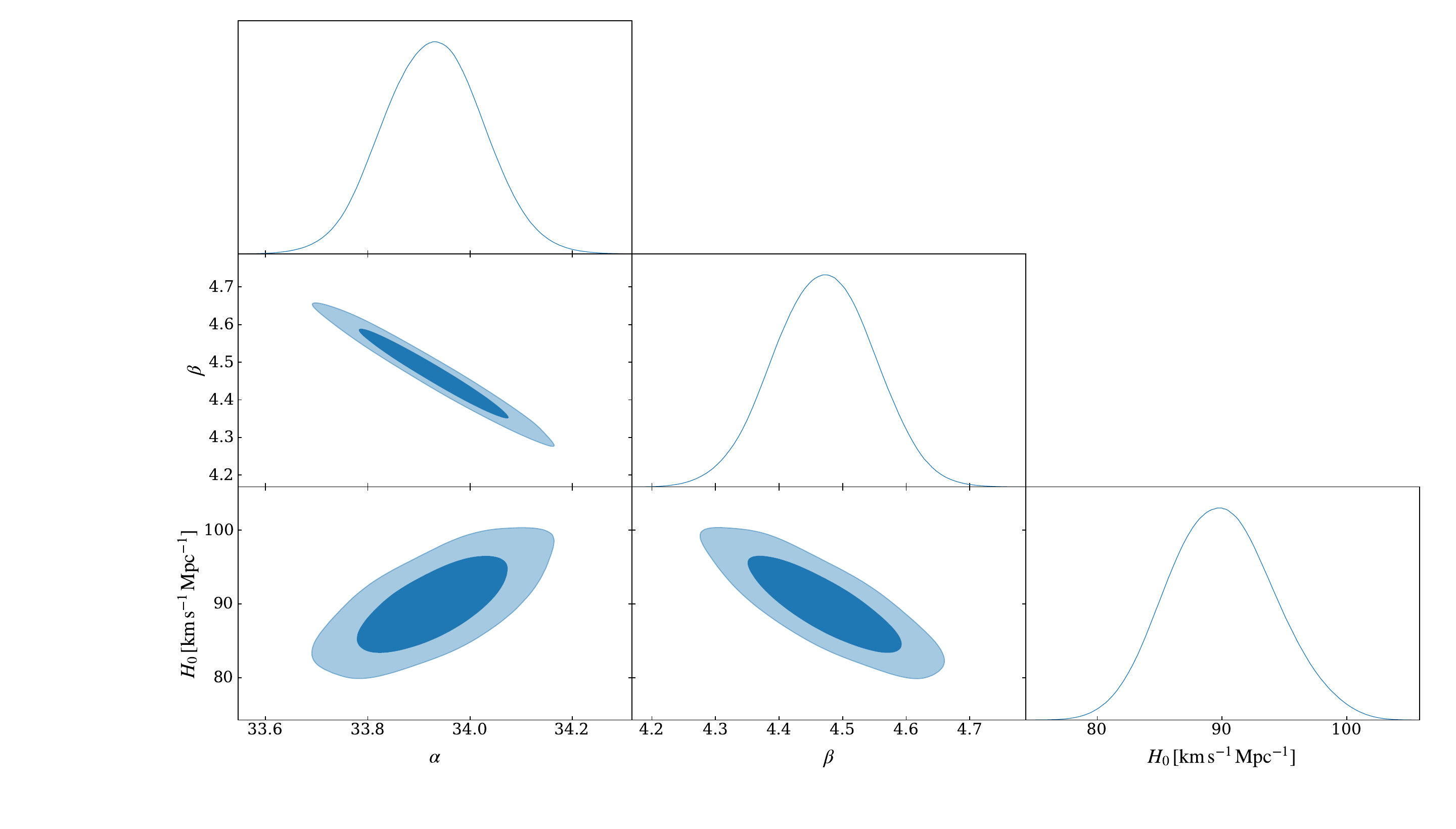}
    \caption{\label{fig:rhct}
  The marginalized 68\% and 95\% credible intervals for the   parameters $\alpha,\beta,H_0$ in $R_\mathrm{h}=ct$.}
\end{figure*}

\begin{figure*}
    \centering
    \includegraphics[width=0.9\textwidth]{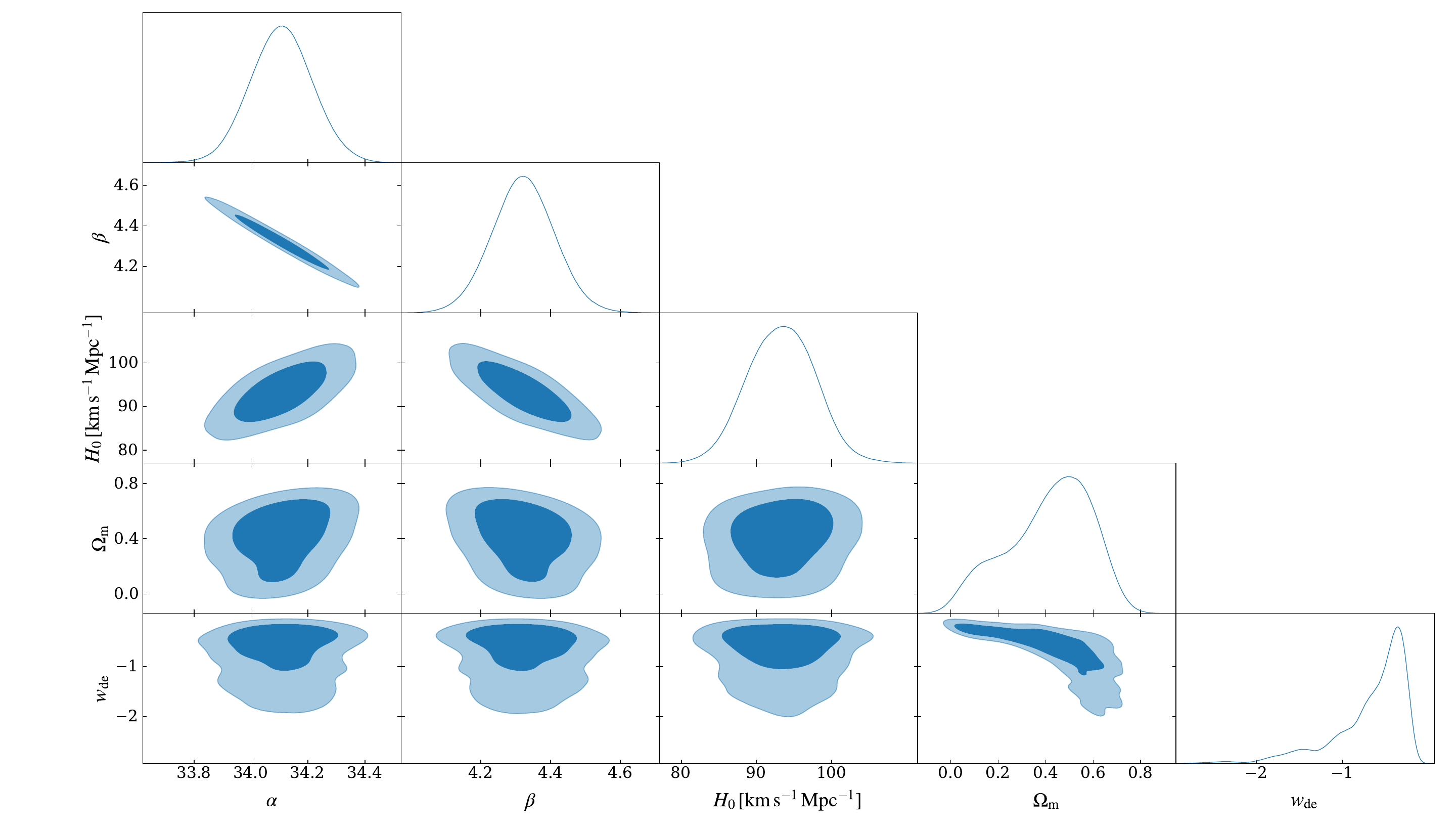}
    \caption{\label{fig:wcdm}
   The marginalized 68\% and 95\% credible intervals for the parameters $\alpha,\beta,H_0,\Omega_m$ and $w_{\mathrm{de}}$ in the (flat) $w$CDM.}
\end{figure*}

As a starting point, we carry out parameter estimation for all three cosmological models after using the same uniform priors on the cosmological parameters used to calculate the Bayesian evidence.  We then sample  the posterior using the  {\tt emcee} based MCMC sampler~\citep{emcee}. The posterior intervals for $\Lambda$CDM, $R_\mathrm{h}=ct$, and 
$w$CDM can be found in Fig.~\ref{fig:lcdm},  Fig.~\ref{fig:rhct}, and 
Fig.~\ref{fig:wcdm}, respectively. We also show the best-fit values (without scatter) for all the parameters in Table~\ref{bestfit}. This table also shows the fractional discrepancy compared to the same values estimated in WM25.
The marginalized 68\% and 95\% credible  intervals  (obtained using {\tt getdist}~\citep{getdist}) for almost all parameters are consistent to within $2\sigma$.

\begin{table*}
\begin{tabular}{ |c|c|c|c|c| } 
\hline
\textbf{Model} & \textbf{Parameter} & \textbf{Prior} & \textbf{Best-fit} & \textbf{Discrepancy with WM25 ($\sigma$)} \\
\hline
\multirow{4}{*}{$\Lambda$CDM} & $\alpha$ & $\mathcal{U} (33,35)$ &  $34.09 \pm 0.11$ & 2.2 \\ 
& $\beta$ & $\mathcal{U} (4,5)$ & 4.34 $\pm$ 0.09 & 2.2 \\ 
& $H_0$ & $\mathcal{U} (60,110)$ & $93.20^{+4.44}_{-4.54}$ & 1.7 \\ 
& $\Omega_\mathrm{m}$ & $\mathcal{U}(0,1)$ & $0.56^{+0.07}_{-0.06}$ & 1.3 \\ 
\hline
\multirow{3}{*}{$R_\mathrm{h}=ct$} & $\alpha$ & $\mathcal{U} (33,35)$ & $33.92^{+0.10}_{-0.09}$ & 1.8 \\ 
& $\beta$ & $\mathcal{U} (4,5)$ & $4.48^{+0.07}_{-0.08}$ & 1.8 \\ 
& $H_0$ & $\mathcal{U} (60,110)$ & $89.87^{+4.39}_{-4.00}$ & 1.5 \\ 
\hline
\multirow{5}{*}{$w$CDM} & $\alpha$ & $\mathcal{U} (33,35)$ & $34.10 \pm 0.11$ & 2.2\\ 
& $\beta$ & $\mathcal{U} (4,5)$ & $4.32 \pm 0.09$ & 2.3 \\ 
& $H_0$ & $\mathcal{U} (60,110)$ & $92.99^{+4.17}_{-4.31}$ & 1.7 \\ 
& $\Omega_\mathrm{m}$ & $\mathcal{U}(0,1)$ & $0.44^{+0.15}_{-0.22}$ & $< 1$ \\ 
& $w_\mathrm{de}$ & $\mathcal{U}(-3,0)$ & $-0.55^{+0.26}_{-0.55}$ & $< 1$ \\ 
\hline
\end{tabular}
\caption{\label{bestfit} Best-fit values for all the cosmological and astrophysical parameters as well as the relative discrepancy compared to WM25 in terms of $Z$-score.  The $Z$-score was obtained based on the ratio of the difference between the two estimates divided by the total error, which was obtained by adding the two errors in quadrature.}
\end{table*}

We now calculate the BIC for all the three cosmological models, with and without the inclusion of intrinsic scatter. In order to calculate $\mathcal{L}_{max}$, we used the maximum value of $\mathcal{L}$ obtained from the MCMC chains. We also cross-checked  this value by maximizing the likelihood using the optimization algorithms available in {\tt scipy}, and the results are comparable. The values of BIC which we get without the intrinsic scatter as well as with intrinsic scatter can be found in Table~\ref{BICvalues} and Table~\ref{BICvalues2}, respectively.
Unlike WM25, we find that the flat  $\Lambda$CDM shows the smallest value for BIC.
The difference in BIC between the flat $\Lambda$CDM and $R_h=ct$ is $\sim 1$ regardless of whether we include scatter or not.  This result disagrees with WM25 (without including scatter), which had found a $\Delta$BIC =5, with $R_\mathrm{h}=ct$ having the smaller value.  \rthis{One possible reason for the disagreement could be due to the $(1-2)\sigma$ discrepancies between our best-fit parameters and those in WM25 (cf. Table~\ref{bestfit}), since BIC only makes use of the best-fit parameters, which maximize the likelihood.}

However, our results for model comparison with BIC  are consistent with those obtained using Bayesian model comparison in the main body of the manuscript. Our results for $\Delta$BIC between $R_\mathrm{h}=ct$ and $w$CDM are  in agreement with WM25.


\begin{table}
\centering
\begin{tabular}{| c | c |}
\hline
\textbf{Model} & \textbf{BIC} \\
\hline
$R_\mathrm{h}=ct$ & 432.7\\
$\Lambda$CDM & 431.9\\
$w$CDM & 435.4\\
\hline
\end{tabular}
\caption{\label{BICvalues} BIC for the models $R_\mathrm{h}=ct$, $\Lambda$CDM and $w$CDM, without accounting for intrinsic scatter (cf. Eq~\ref{eq:epsanchor})}
\end{table}

\begin{table}[h]
\centering
\begin{tabular}{|c|c|}
\hline
\textbf{Model} & \textbf{BIC} \\
\hline
$R_\mathrm{h}=ct$ & 277.0\\
$\Lambda$CDM & 277.0\\
$w$CDM & 282.0\\
\hline
\end{tabular}
\caption{\label{BICvalues2} BIC for the models $R_\mathrm{h}=ct$, $\Lambda$CDM and $w$CDM, after accounting for intrinsic scatter (cf. Eq.~\ref{eq:epshiimod}).}
\end{table}

\section{Sensitivity of BF to different prior choices}
\rthis{One reason we chose very broad uniform priors especially on $\Omega_m$ and $w$ while calculating the Bayes factors is so that we can get closed contours for the posteriors. Although we have already chosen normal priors based on Planck 2020 cosmology, in order to test the robustness of our conclusions we also redo our analysis with narrow range of uniform priors. We choose $\Omega_m \in [0.2,0.4]$  for both $\Lambda$CDM and $w$CDM and $w \in [-2,0]$ for $w$CDM. The priors on $H_0$ are same as before. Our results with the new sets of priors can be found in Table~\ref{BF3}. We find  that although there are minor changes  in the Bayes factors  compared to those calculated in Table~\ref{tablemodbf1}, none of the Bayes factors exceed 50 to justify a strong preference for $R_h=ct$ over $\Lambda$CDM/$w$CDM.}

\begin{table}
\centering
\begin{tabular}{|c|c|c|c|} \hline
\textbf{Prior} & \textbf{Model}  & $\ln (Z)$  & \textbf{BF}[$(R_\mathrm{h}=ct)/(\Lambda \mid w\mathrm{CDM})$] \\ \hline
$\Omega_m \in [0.2,0.4]$ & $\Lambda$CDM & $-219.07 \pm 0.07$ & $15 \pm 1.5$ \\
$\Omega_m \in [0.2,0.4]$; $w \in [-2,0]$ & $w$CDM & $-214.40 \pm 0.07$ & $0.14 \pm 0.01$ \\
$\Omega_m \in [0,1]$; $w \in [-2,0]$ & $w$CDM & $-214.62 \pm 0.07$ & $0.18 \pm 0.02$ \\
\hline
\end{tabular}
\caption{\label{BF3}\rthis{Sensitivity of Bayes factors to different choices of priors. The first column indicates the choice of priors used. For this table we have used $H_0 \in [60,110]$ km/sec/Mpc. The third column gives the natural log of Bayesian evidence for $\Lambda$CDM or wCDM and the last column gives the Bayes factor for $R_\mathrm{h}=ct$ compared to $\Lambda$CDM or $w$CDM using the Bayesian evidence for $R_\mathrm{h}=ct$ calculated in Table~\ref{tablemodbf1} for a uniform prior in $H_0$.}}
\end{table}
\end{document}